OPEN

# A pathway to peptides in space through the condensation of atomic carbon

S. A. Krasnokutski[1]✉, K.-J. Chuang[1,4], C. Jäger[1], N. Ueberschaar[2] and Th. Henning[3]

**Organic molecules are widely present in the dense interstellar medium, and many have been synthesized in the laboratory on Earth under the conditions typical for an interstellar environment. Until now, however, only relatively small molecules of biological interest have been demonstrated to form experimentally under typical space conditions. Here we prove experimentally that the condensation of carbon atoms on the surface of cold solid particles (cosmic dust) leads to the formation of isomeric polyglycine monomers (aminoketene molecules). Following encounters between aminoketene molecules, they polymerize to produce peptides of different lengths. The chemistry involves three of the most abundant species (CO, C and $NH_3$) present in star-forming molecular clouds, and proceeds via a novel pathway that skips the stage of amino acid formation in protein synthesis. The process is efficient, even at low temperatures, without irradiation or the presence of water. The delivery of biopolymers formed by this chemistry to rocky planets in the habitable zone might be an important element in the origins of life.**

The origin of life has always been one of the most intriguing questions throughout human history. Biomolecules delivered to early Earth by asteroids, meteorites or comets during the period of heavy bombardment about four billion years ago have been proposed to play a role in the origin of life[1,2]. Similar processes may apply to rocky exoplanets (or their moons). Analysis of meteoritic material led to the identification of amino acids, sugars and nucleobases, among other complex organic molecules of extraterrestrial origin[3,4]. The simplest amino acid, glycine, has been discovered in comets[5]. The widespread hypothesis of the formation of organic molecules in space suggests that they are synthesized in the icy mantle that covers the refractory particles of cosmic dust[6,7]. At later stages, these dust grains form asteroids and comets. These bodies may have a liquid phase at low temperatures, where water mixes with gases such as $CO_2$ and $NH_3$ (refs. [8–10]). The chemistry in the liquid phase may further enhance the molecular complexity. Experimentally, the formation of various amino acids, and even their dimers (as well as other organic molecules) has been detected following energetic processing of different molecular ices[11,12]. More recently, non-energetic pathways were also found for small organic molecules[13,14] and glycine[15,16]. However, the formation of biopolymers under space conditions has not been demonstrated, although several chemical pathways in terrestrial conditions have been widely accepted[17–19].

It is generally assumed that the prebiotic synthesis of peptides occurs in two stages. The first stage is the formation of amino acids, and the second is their polymerization[20]. The polymerization process requires the condensation of amino acids accompanied by the loss of water. This has a high energy barrier and therefore proceeds only at high temperatures or requires energetic processing of the material[21,22]. Consequently, each of these stages has a relatively low probability. Instead of first synthesizing amino acids to subsequently dehydrate them for the polymerization process, we suggest a much simpler and direct formation of the aminoketene ($NH_2CH{=}C{=}O$) and its polymerization forming peptides under astrophysically relevant conditions. Quantum chemical calculations predict that the $CO + C + NH_3 \xrightarrow{\text{surface}} NH_2CH{=}C{=}O$ reaction is barrierless, and therefore occurs on the cold dust grains without involving external energy[23]. Although isolated aminoketene is theoretically predicted to be stable[24,25], it has never been observed in experimental studies to the best of our knowledge. This can be related to the extremely high reactivity of the aminoketene, which could also lead to its efficient polymerization. The polymerization of $NH_2CH{=}C{=}O$ via the nucleophilic attack of the nitrogen on the carbonyl carbon and an intramolecular proton transfer from $NH_2$ to CH results in the formation of peptide chains. Therefore, in contrast to the polymerization of amino acids, the formation of polyglycine via polymerization of aminoketene is a much simpler process.

To test this reaction pathway experimentally, we performed the co-deposition of CO, C and $NH_3$ on the surface of both Si and KBr substrates cooled to 10 K and placed inside an ultrahigh vacuum (UHV) chamber to mimic the conditions in dense molecular clouds.

## Chemistry at 10 K

Infrared (IR) absorption spectra of the ice produced after the co-deposition of $CO + NH_3$ and $CO + C + NH_3$ at low temperature ($T = 10$ K) are shown in Fig. 1. The additional control experiments, involving only two reactants C + CO and $C + NH_3$, are shown in Extended Data Figs. 1 and 2.

The IR spectrum representing the deposition of all three reactants (CO, C and $NH_3$) on the substrate at 10 K reveals several new absorption features, which are absent in reference spectra involving only two reactants (for example, C + CO, $C + NH_3$ and $CO + NH_3$). Therefore, these bands have to be assigned to a product formed by reactions of all three reactants at 10 K. In addition, the warming up of the CO/C/$NH_3$ ice layer to room temperature leaves a residue on the substrate, which is also observed for the reaction of C atoms with only CO or $NH_3$, but in much smaller amounts. We conclude that the product formed in the reaction of all three reactants at 10 K is required for the formation of the non-volatile products at 300 K. As C atoms are the limiting reactant (that is, the fluence of C atoms was at least ten times less than that of $NH_3$ or CO), the reactions of

[1]Laboratory Astrophysics Group of the Max Planck Institute for Astronomy, Friedrich Schiller University Jena, Jena, Germany. [2]Mass Spectrometry Platform, Faculty of Chemistry and Earth Sciences, Friedrich Schiller University Jena, Jena, Germany. [3]Max Planck Institute for Astronomy, Heidelberg, Germany. [4]Present address: Leiden Observatory, Leiden University, Leiden, the Netherlands. ✉e-mail: sergiy.krasnokutskiy@uni-jena.de

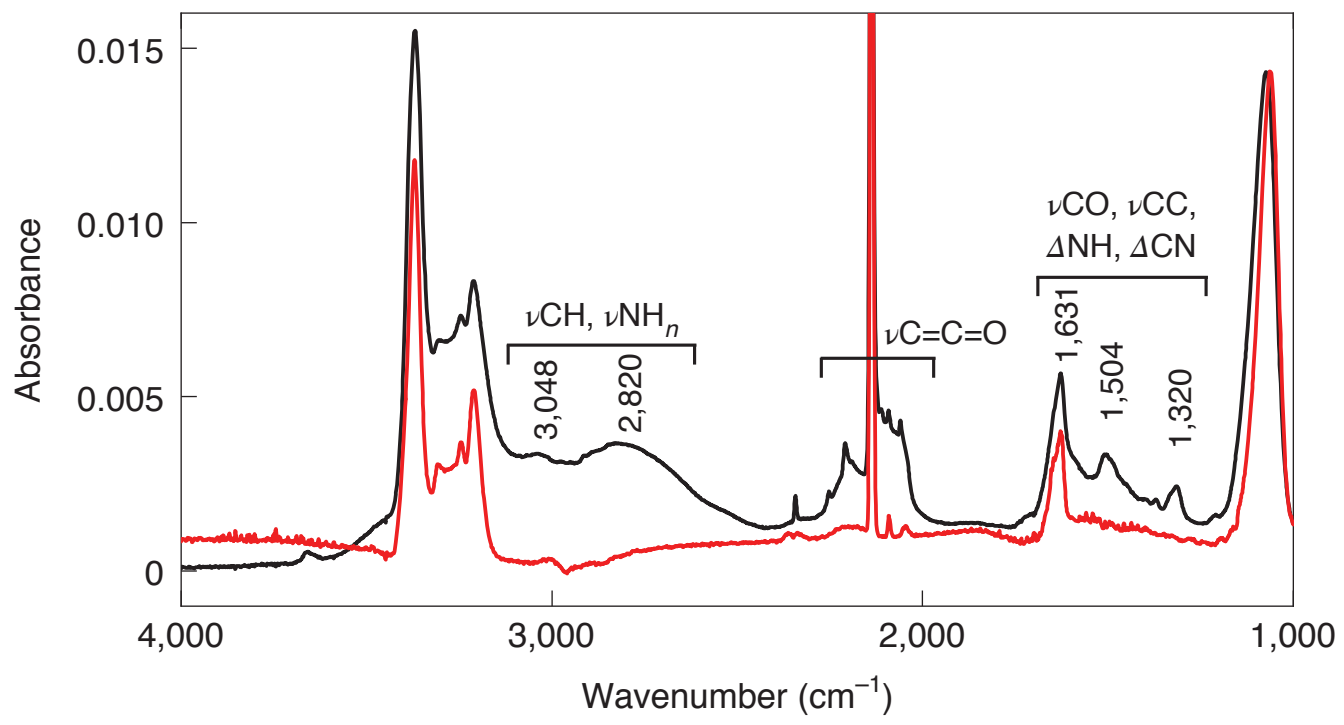

**Fig. 1 | IR absorption spectra of the ice mixtures produced at 10 K.** The red line shows the IR spectrum of $CO/NH_3$ (1:1) ice and the black line the spectrum of $CO/NH_3/C$ (1:1:0.1) ice. Stretching and bending vibrations are denoted $\nu$ and $\Delta$, respectively.

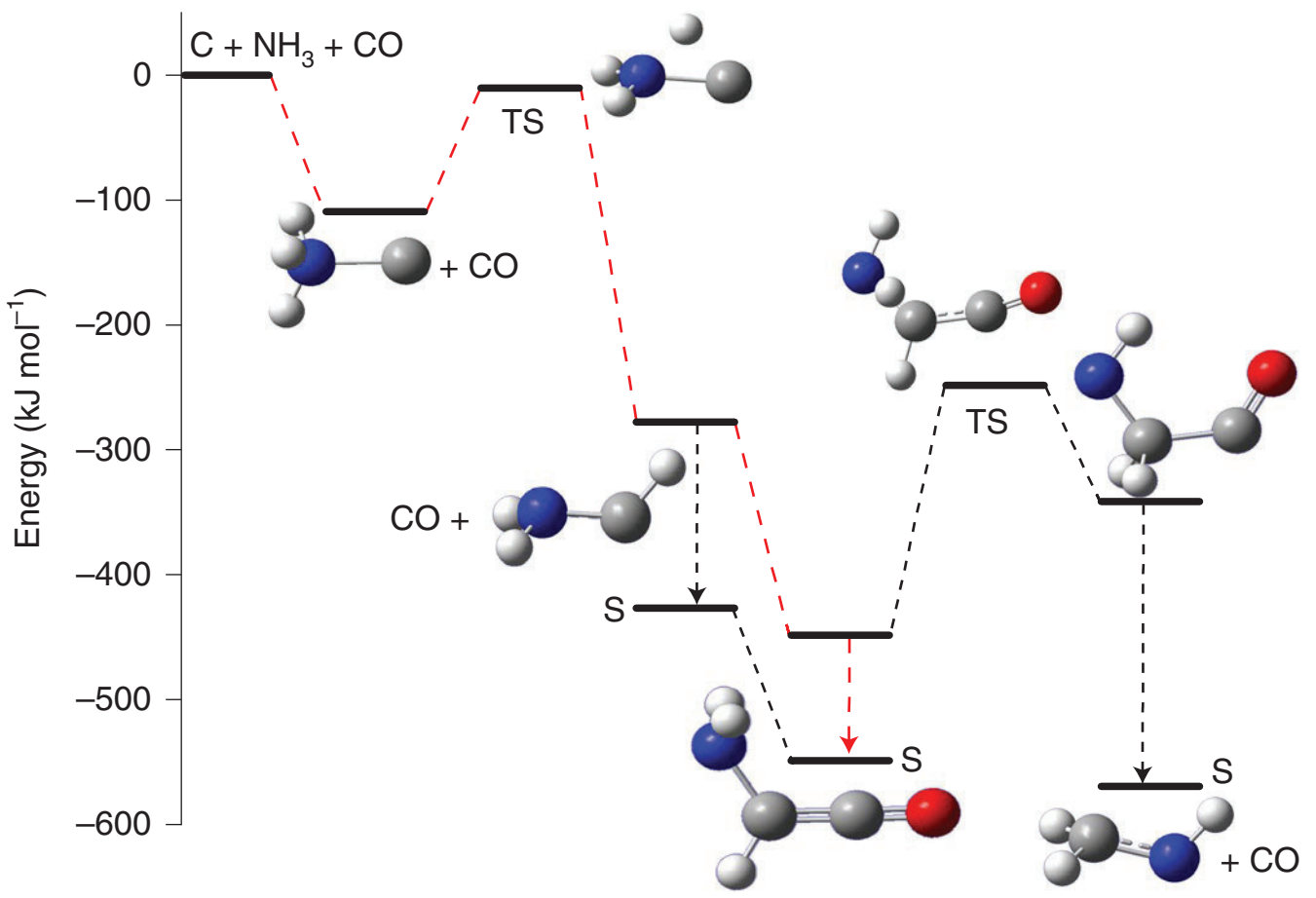

**Fig. 2 | Energy level diagram for the reaction involving CO, C and $NH_3$ reactants.** The reaction starts with the triplet state. The transition states (TS) and the singlet states (S) are labelled and the dashed lines show the possible reaction pathways, with the most probable pathway based on computational and experimental results in red.

C atoms with other C atoms or with its reaction products are negligible. Consequently, the observed product is expected to originate from the reaction of $CO+C+NH_3$ and contains CH, CN, C=O and $NH_n$ groups, as shown in the IR spectrum (black line in Fig. 1). Given that $NH_3$ does not react with CO at 10 K, the addition of C atoms must act as a chemical trigger to initiate such solid-state reactions. The formation of the C=C=O functional group is demonstrated by the appearance of the absorption bands in the ~1,965–2,275 cm$^{-1}$ range. It implies the transformation of the CO triple bond (C≡O) to a CO double bond (C=O) species by attaching a C atom to the C atom of carbon monoxide. Similarly, the detection of CH and $NH_n$ (2,500–3,000 cm$^{-1}$) and CN (1,320–1,500 cm$^{-1}$) absorption bands also suggests covalent bond formation between the added C atom and ammonia: C-atom addition to N atoms of $NH_3$ followed by H-atom rearrangement. On the basis of the above considerations, we can conclude that $O{=}C{=}CH_mNH_n$ molecules are formed.

To better understand the chemistry involving all three reactants at 10 K, we performed quantum chemical calculations (results are shown in Fig. 2 and Extended Data Fig. 3). As shown in Extended Data Fig. 3, C atoms initially prefer to react with ammonia molecules, forming the prereactive complex $CO+CNH_3$, which corresponds to the first energy well in Fig. 2. The $C+NH_3$ reaction pathway has been studied experimentally[16] using the recently developed calorimetric technique[26]. The proton transfer from nitrogen to carbon was observed to form $NH_2CH$ before the energy dissipation[16]. This mechanism also applies to the reaction investigated here, as shown by the appearance of $NH_n$ and CH absorption bands (~3,048 and ~2,820 cm$^{-1}$, Fig. 1) after C-atom addition. The formed $NH_2CH$ product subsequently reacts barrierlessly with CO, leading to the formation of $NH_2CH{=}CO$. There are thus two possible products of this reaction: $NH_2CH{=}C{=}O$ and $CH_2NH+CO$. The latter is the same as the product of the reaction with only two reactants $C+NH_3 \rightarrow CH_2NH$. Therefore, the reaction product observed upon the addition of all three reactants at 10 K can only be $NH_2CH{=}C{=}O$, the functional groups of which are observed in the IR spectrum. The formation of this molecule was confirmed by temperature-programmed desorption in combination with quadrupole mass spectrometer analyses monitoring the ions with masses of 57 and 56 unified atomic mass units (u) as shown in Extended Data Fig. 4. At the same time, C atoms could react with either CO or $NH_3$, and a smaller number of CCO or $H_2CNH$, or even $NH_2CH_2NH_2$, molecules can also be formed.

## Chemical transformation during temperature rise

After deposition, the substrate was heated at a rate of 2 K min$^{-1}$. The IR spectra measured at selected temperatures are shown in Fig. 3. A comparison of the spectra at 10 K and 70 K shows that the sublimation of CO, occurring mainly from 25 to 40 K, does not initiate a new chemistry. However, with the sublimation of ammonia (above 110 K), the products of the C-atom reactions can finally encounter each other and react. We observed the formation and growth of new IR absorption bands, while the intensity of the $\nu NH_n$ and $\nu$C=O bands as shown in Fig. 3 decreased significantly. With the mass spectrometer, we did not detect any considerable desorption from the substrate besides CO and $NH_3$. Therefore, the disappearance of these IR bands was due to chemical transformation rather than sublimation. The decrease of band intensities in the 2,800–3,050 cm$^{-1}$ range is a common indicator of glycine polymerization[27]. This occurs due to the transformation of the $NH_3^+$ or $NH_2$ groups of glycine into the NH groups present in peptides. Simultaneously with the transformation of $NH_n$ groups, we observed the formation of the peptide I band at 1,670 cm$^{-1}$ and peptide II band with a peak maximum ranging from 1,548 to 1,569 cm$^{-1}$ (Fig. 3 and Extended Data Fig. 5), which are the characteristic features of peptide bonds. The peptide bond is therefore formed at 100–120 K simultaneously with the decrease in the intensity of the IR bands associated with the molecules present in the ice at 10 K (Extended Data Fig. 6). The IR spectrum did not change substantially above 150 K; the residual material present on the substrate after warming up to 300 K (R300K) must therefore have formed at low temperature during ammonia sublimation.

## Characterization of the room-temperature residue

The in situ characterization was performed by IR spectroscopy. After removing the substrate from the UHV chamber and exposing it to air, we did not observe any noticeable modification of the IR spectrum of R300K. Therefore, an ex situ analysis could also be performed. The high-resolution ex situ mass spectrum of R300K is shown in Fig. 4. To better characterize the mass spectrum, we selected a threshold shown by the horizontal dashed line in Fig. 4. The peaks with intensities higher than this threshold were analysed to determine the series. We could identify many series of mass peaks, which are separated by the same mass of about 57.0215 u. This value exactly matches the mass of the $NH_2CH{=}C{=}O$ molecule, which was found to be the main product formed at 10 K. The intensities of the peaks in these series drop exponentially with increasing mass. A series is formed if the same molecule is continuously added to an initial species. Therefore, the presence of these series with such an intensity distribution unambiguously indicates the polymer formation with the mass of the monomer unit equal to the distance between the peaks. The complete list of all series found is given in Extended Data Fig. 7. As shown in Fig. 4, most of the analysed mass peaks belong to the series.

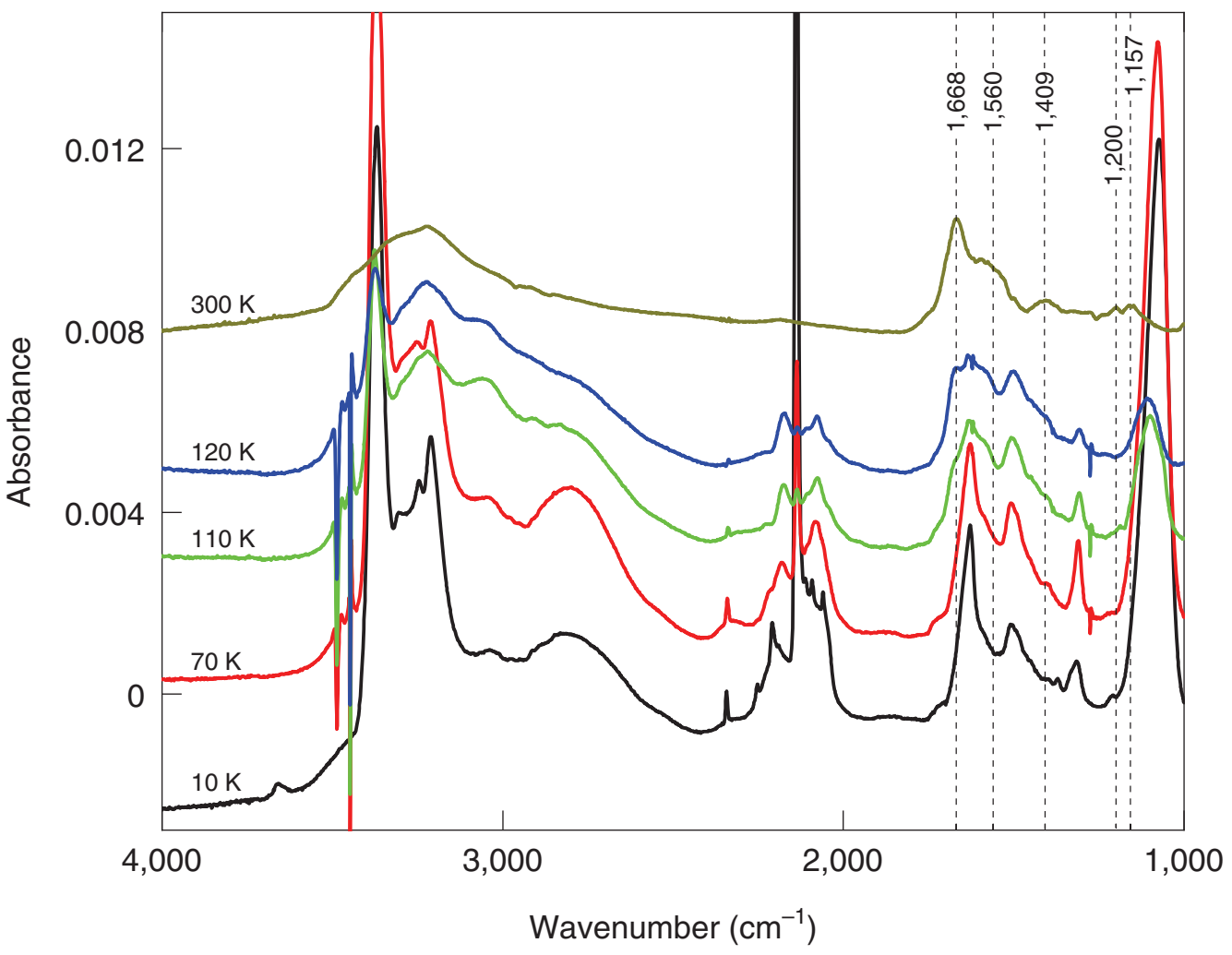

**Fig. 3 | IR absorption spectra obtained during annealing of the material produced by co-deposition of CO + C + $NH_3$.** The labels for the vertical dashed lines indicate the positions of the peaks appearing during heating at the temperatures indicated. All CO is sublimated at 70 K and most $NH_3$ is sublimated at 120 K.

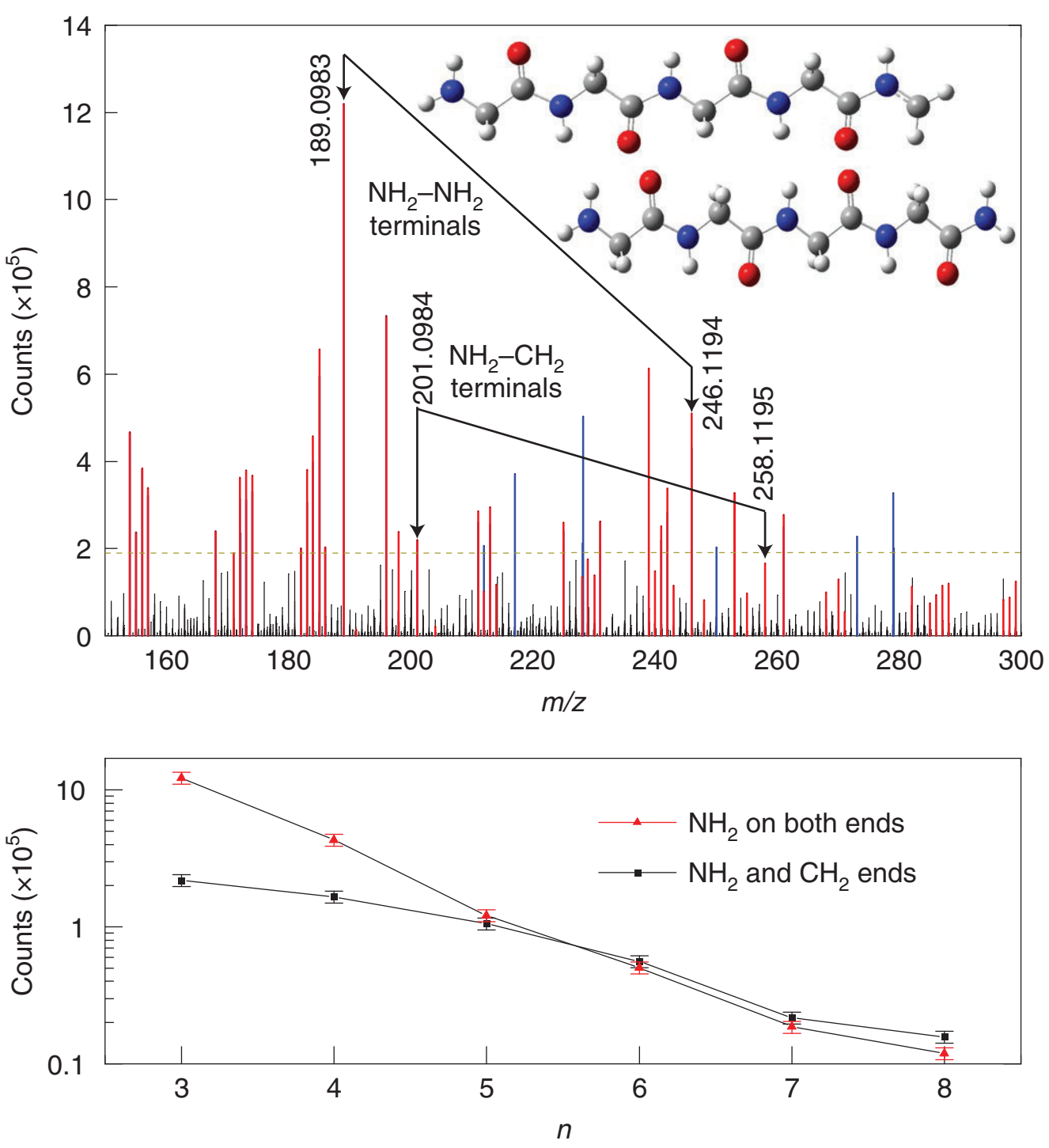

**Fig. 4 | Ex situ mass spectrometric analysis of the R300K residue.** Top: the mass spectra of R300K obtained from ex situ analysis using an ESI-Orbitrap mass spectrometer. The horizontal orange dashed line shows the threshold used to select the analysed peaks. The peaks that belong (or do not belong) to the series are displayed in red (blue). Black peaks lie below threshold and were not analysed for series. The molecular structures display the $GI_4$ peptides detected in the mass spectra as protonated species [analyte + H]$^+$. Bottom: the intensities of the bands derived from two observed series corresponding to glycine peptides are plotted as a function of the numbers of monomeric units (*n*). Error bars are standard deviations of corresponding peak intensities in the measured mass spectra.

We can therefore conclude that the polymerization of the $NH_2CH{=}C{=}O$ molecule formed at 10 K, is the main chemical pathway leading to the formation of R300K. The formation of a variety of different series (Extended Data Fig. 7) can be understood assuming that other molecules formed at 10 K could be added during the polymer formation.

The IR spectrum of R300K is shown in Fig. 5 (see also details in Extended Data Fig. 5). It highlights the absorption features characteristic for the peptide bonds. As the $NH_2CH{=}C{=}O$ molecule does not contain the peptide bond, the appearance of this bond in the formed polymer indicates the type of the bond between the monomeric units. Upon sublimation of volatile species, the monomeric units of $NH_2CH{=}C{=}O$ are assembled through peptide bonds, resulting in glycine oligomers, which differ from canonical peptides only by the groups at their C termini. This proceeds by the polymerization of $NH_2CH{=}C{=}O$ molecules, involving proton transfer from nitrogen to the carbon atom of the CH group. However, our density functional theory calculations show that the $NHCH_2C{=}O$ (as well as $NH_2CHCO{-}NHCH_2C{=}O$) molecules are not stable; they are expected to fragment with the loss of the CO molecule. Therefore, peptide chains produced by this mechanism are terminated by $NH_2$ on one side, which corresponds to the canonical structure of the peptide. The other side of the peptide chain is terminated by a $CH_2$ group instead of a classical COOH group. However, the observed series is not the most intense one. The most intense peak (189.0983 u) shown in Fig. 4 suggests that the favoured oligopeptides are terminated by amino groups on both sides, which assumes an efficient involvement of ammonia in the polymerization process. In spite of the differences in length and terminal functional groups between the canonical $GI_3$ peptide and the peptides identified in our experiment, there is a good match between the two spectra shown in Fig. 5. The main discrepancy is the width of the IR absorption bands; this is expected owing to the formation of oligopeptides with a wide distribution of lengths. The band broadening is caused by the change in the positions of the IR absorption bands with the number of units in the glycine oligomers[28].

Finally, to confirm the proposed molecular structure of the polymers, we performed the higher-energy C-trap dissociation[29] of the most abundant ion from Fig. 4. The result of this measurement is shown in Fig. 6. This analysis explicitly confirms the suggested molecular structure of this ion with mass 189.0984 u. The efficient loss of two ammonia molecules demonstrated by peaks b3 and b3* clearly shows the terminal location of both amino groups. The a3 peak shows that the side amino group is linked with the CO unit. The b2 and y1 peaks demonstrate that the side $H_2NCO$ unit is connected to the $CH_2NH$ unit. The same analysis can also be performed from the other side, where the terminal amino group is joined to the $CH_2CO$ group, as demonstrated by the y2 fragmentation peak. The fragmented ion (189.0983 u) belongs to the series of peaks that considerably exceeds the intensities of other series. Therefore, the mass peaks from this series can be separated by both mass and intensity. This shows that the formation of the ions from this series happens via polymerization of the $NH_2CH{=}C{=}O$ molecule. All of this unambiguously confirms the suggested molecular structures of the ions from this series. Moreover, it also strongly suggests that ions from other series shown in Fig. 4 have a similar molecular structure and are modified peptides. We evaluated the entire mass spectrum, comparing the integrated intensities of mass peaks in series and the total integrated intensities of all analysed peaks. This analysis determined a polyglycine concentration in our sample of about 85%, given the same ratio between the polyglycine peaks and other mass peaks among non-analysed peaks below the threshold.

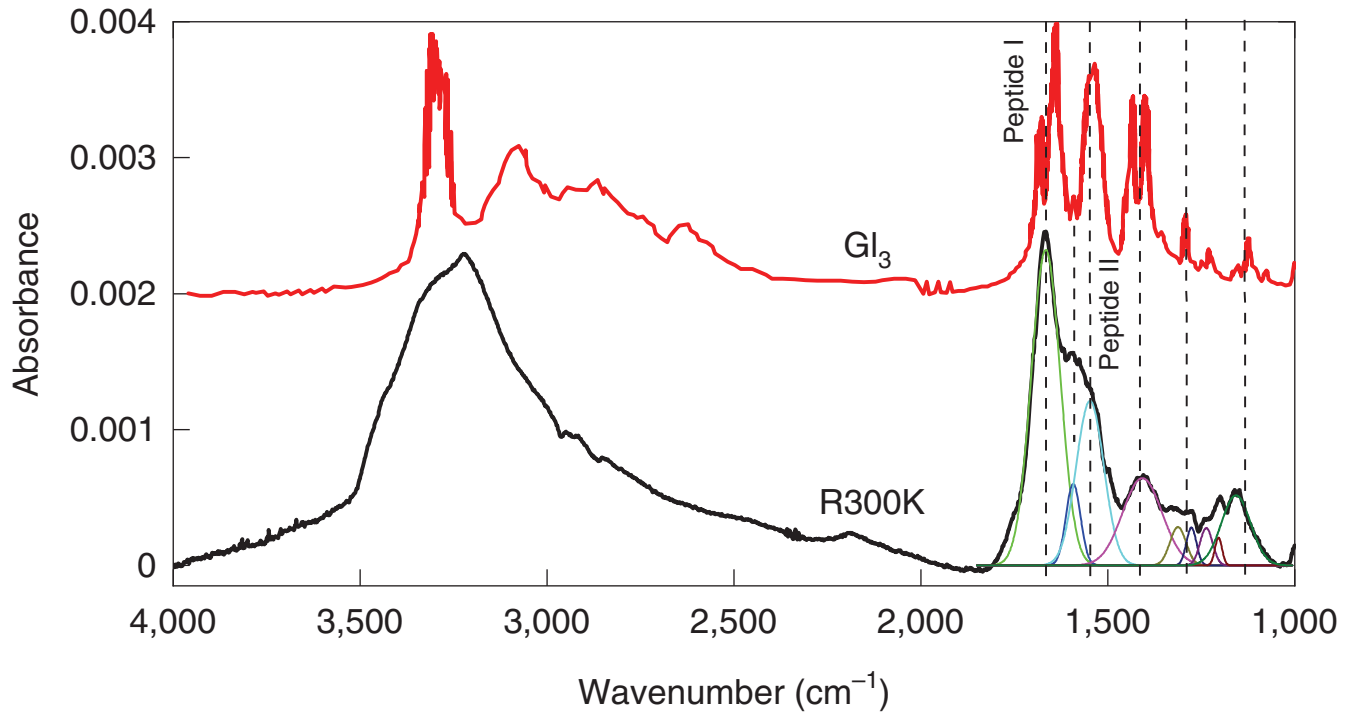

**Fig. 5 | Comparison between the IR absorption spectra obtained from R300K and $GI_3$.** The vertical dashed lines are drawn to visualize the coincidences between the peaks in the two spectra. The Gaussian profiles beneath R300K spectrum show the deconvolution of the experimental spectrum (Extended Data Fig. 5). The spectrum of triglycine (Gl3) is adopted from ref. [28].

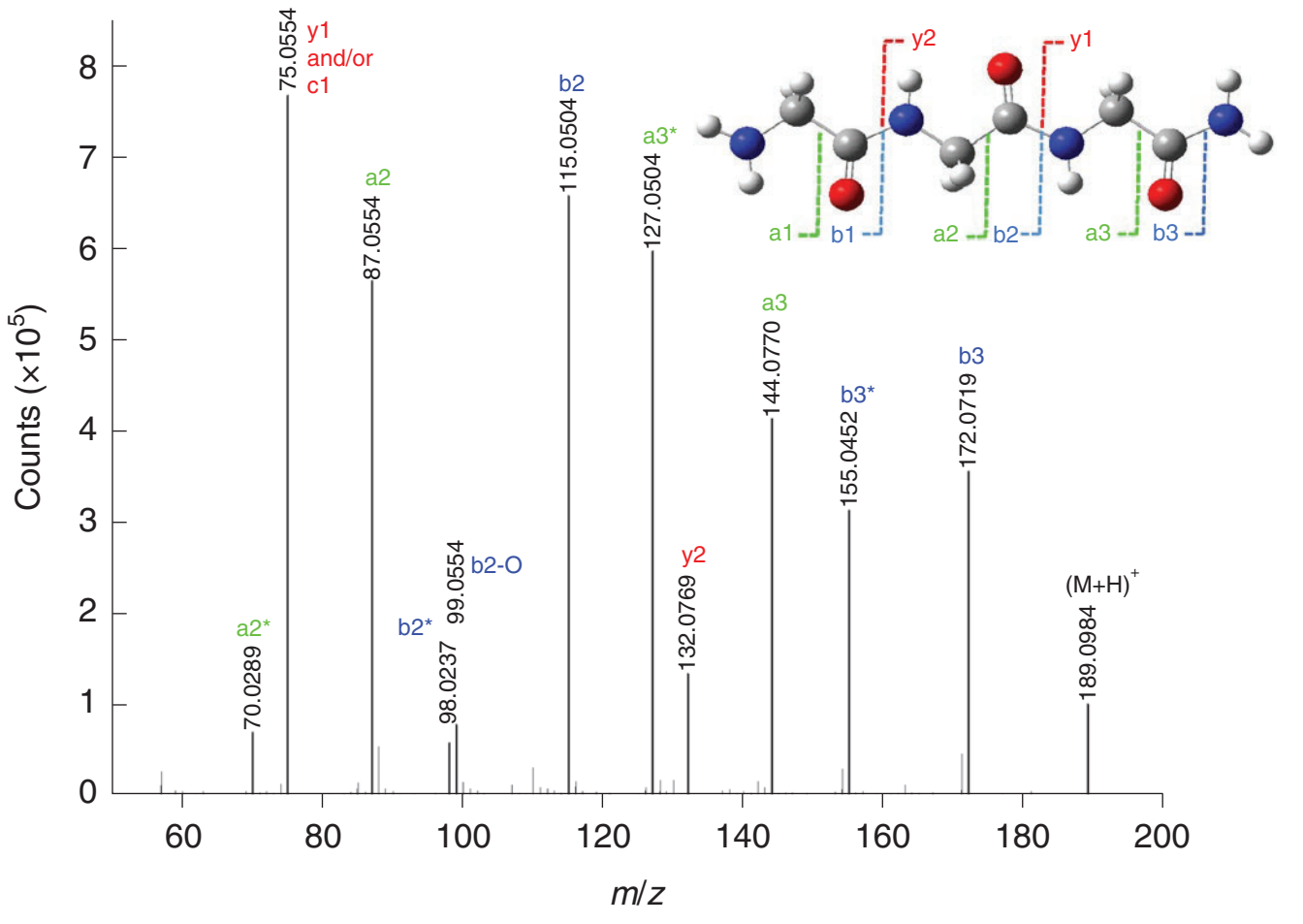

**Fig. 6 | The higher-energy C-trap dissociation of the 189.0983 u ion.** The fragmentation peaks are annotated based on the accepted nomenclature for the peptide fragments[29,44]. The loss of an additional ammonia molecule is indicated with an asterisk. The inset shows the molecular structure of the investigated ion. The coloured marks (a, b, y) determine the fragments that correspond to the mass peaks identified in the spectrum.

If all non-analysed (weak) peaks did not belong to polyglycine, the value would be reduced to 19%. Considering that the R300K material is completely solvated in water and that the ratio of the intensities of the peptide bands towards other IR bands in the spectrum of our sample is comparable to that in the spectrum of pure oligoglycine, we concluded that the predominantly formed molecules in our experiments are polyglycine.

## Implications

The reactants used in this work are among the most abundant species present in the interstellar medium (ISM). The fractional abundances of $NH_3$ and CO in the ice mantles covering refractory dust particles are 10% and 40%, respectively[30]. In the ISM, more than half of all carbon exists in the form of atomic gas[31]. The formation of dense clouds, where new stars and planets are formed, goes through a stage of translucent clouds, where a notable portion of carbonaceous dust is expected to be formed due to the accretion of C atoms. In translucent molecular clouds, the dust temperature is low enough to allow the formation of molecular ices while carbon is dominantly present in the atomic or CO forms[32]. Therefore, reactions between accreted CO, C and $NH_3$ could be very common in this region. At the low temperatures of dust in translucent molecular clouds ($T$=10–20 K)[33], this leads to the formation of organics[34]. As demonstrated here, a portion of these organics could be in the form of peptides. At later stages, such dust becomes the building blocks of comets or meteorites. The formed organics could therefore have been delivered to the early Earth during the period of heavy bombardment. The survival of organics at all delivery stages is thus a relevant question. In this sense, comets have the advantage of being less mechanically stable than asteroids, because their fragmentation after entering the atmosphere may allow strong impacts to be avoided due to the loss of kinetic energy in the atmosphere as the meteor breaks up[2]. Comets might deliver intact organics produced in the ISM at a rate of at least $10^6$–$10^7$ kg yr$^{-1}$ (ref. [2]). However, even the direct impact of larger objects may allow the partial survival of molecules, including amino acids[35,36]. The biggest issue with this extraterrestrial scenario is the stability of the newly formed organics in the harsh energetic environment of the ISM, filled with X-ray and ultraviolet photons. The non-energetic formation of peptides may be advantageous in this regard: if gas-to-solid phase transitions and reactions occur in regions with a low ultraviolet flux, the formed peptides could survive long enough to be incorporated into bulk solids, which would shield them from further destruction. This is in line with the potential detection of proteins in meteorites[37,38]. Peptides also play a key role in the origin of life[20]. The delivery of peptides to Earth should therefore expand the opportunities for evolutionary chemistry and biology. Moreover, the peptides with two amino terminals dominantly formed in our experiments are capable of efficient self-assembly[39], which offers the prospect of new and interesting opportunities for abiogenesis.

## Methods

**In situ experiments.** The experiments were performed using the UHV INterStellar Ice Dust Experiment (INSIDE) set-up described elsewhere[40]. We performed the co-deposition of CO, C and $NH_3$ on the surface of KBr substrates cooled down to 10 K and centred in the UHV chamber. The deposition of reactants was performed for about 1 h. Low-energy carbon atoms were generated by an atomic carbon source[41]. The source generated a pure flux of low-energy carbon atoms in the triplet ground state $C(^3P_J)$ and about 1% of $CO/CO_2$ molecules. The background pressure inside the vacuum chamber ($1\times10^{-10}$ mbar) and the temperature of the substrate (10 K) allowed us to mimic the chemistry under dense molecular cloud conditions. We used approximately equal amounts of CO and $NH_3$ molecules, while the number of C atoms was at least ten times smaller. The flux of C atoms from the source was estimated in previous work[42]. This small amount of C atoms allows us to exclude their reactions with each other, as well as with their reaction products. The gases were introduced through two separated leak valves. The ice thicknesses on substrates were monitored by infrared (IR) spectroscopy using a Fourier transform IR spectrometer (Vertex 80 v, Bruker) in the transmission mode. In situ mass spectrometry was performed using a quadrupole mass spectrometer (HXT300M, Hositrad) attached to the same UHV chamber. After the deposition, the substrate was heated at a rate of 2 K min$^{-1}$. As the material was warmed, IR absorption spectra and mass spectra were measured to monitor the residue and sublimated gas-phase species, respectively. The evolution of IR spectra during temperature rise (shown in Extended Data Fig. 6) was obtained by integrating the corresponding bands in the IR spectra.

**Ex situ mass spectrometry analysis.** For the mass spectrometry analysis, we used silicon substrates and performed a 4 h deposition of the reactants CO, C and $NH_3$ under the same conditions as the in situ experiments. We obtained very similar IR spectra at all stages of the experiment, using both Si and KBr substrates. After warming up, the substrates were removed and analysed. The residue was extracted with a water–methanol mixture (70:30) and used for the mass spectral analysis. Complete solubility of the residue in water and water–methanol mixtures was observed. There was also no chemical reactivity of the residue with water, as shown by the unchanged IR spectrum of the material after prolonged exposure to wet air. The mass spectrometry was performed using the hybrid linear trap/Orbitrap mass spectrometer (Thermo Fisher QExactive plus mass spectrometer with a heated ESI source). The accuracy of the mass determination was higher than 5 ppm, which resulted in a possible error only in the fourth decimal place for the investigated mass range. Considering the limited number of elements used in the experiments (C, O, N and H), this high resolution allowed us to unambiguously determine the elemental composition of the detected ions in the mass range shown in Fig. 4. To ensure that the observed mass peaks were generated by the material formed in our experiment, we checked for the possible presence of contaminants (both on the substrate and in the mass spectrometer) using the same procedure with clean silicon substrates.

The same mass spectrometer was used for tandem mass spectrometry measurements. For this purpose, the quadrupole filter was set to preselect ions with the mass range of 188.9–189.3 u (the most abundant $Gl_3$ ion observed in our mass spectra). This cation was then a subject for a higher-energy C-trap dissociation procedure, as described in ref. [29]. The collisional energy was set to 30 eV. In test measurements with 10 and 50 eV collisional energies, we did not observe any principal difference in the fragmentation pattern. Two impurity peaks (189.0734 and 97.0373 u) were subtracted from measured spectra. These peaks were assigned to the impurities on the basis of the presence of this ion signal during the tuning of the collisional energies when all other mass peaks were missing, and the fact that ions with these masses cannot simply be fragments of the parent ion.

**Quantum chemical computations.** Quantum chemical calculations were performed using the GAUSSIAN16 software[43]. To find out the outcome of the reaction of C atoms landing on $CO/NH_3$ ice, we performed a two-dimensional potential energy scan of the $CO + C + NH_3$ reaction at the MP2/6-311+G (*d*, *p*) level. In this scan, we varied the C–C and C–N distances and allowed complete optimization of molecular geometries without any further restrictions. This computation allowed us to identify the formation of the first prereactive complex used for the calculation of the energy level diagram, which is shown in Fig. 2. The geometries of the molecules in the energy level diagram were determined at the B3LYP/6-311+G (*d*, *p*) level. The reaction energies were determined from the difference between the sum of the energies of the reactants and the energy of the product molecules with vibrational zero-point energy corrections.

## Data availability

Source data are provided with this paper.

## References

1. Pearce, B. K. D., Pudritz, R. E., Semenov, D. A. & Henning, T. K. Origin of the RNA world: the fate of nucleobases in warm little ponds. *Proc. Natl Acad. Sci. USA* **114**, 11327–11332 (2017).
2. Chyba, C., Thomas, P., Brookshaw, L. & Sagan, C. Cometary delivery of organic molecules to the early Earth. *Science* **249**, 366–373 (1990).
3. Pizzarello, S. & Cronin, J. R. Alanine enantiomers in the Murchison meteorite. *Nature* **394**, 236–236 (1998).
4. Glavin, D. P., Burton, A. S., Elsila, J. E., Aponte, J. C. & Dworkin, J. P. The search for chiral asymmetry as a potential biosignature in our Solar System. *Chem. Rev.* **120**, 4660–4689 (2020).
5. Altwegg, K. et al. Prebiotic chemicals—amino acid and phosphorus—in the coma of comet 67P/Churyumov-Gerasimenko. *Sci. Adv.* **2**, e1600285 (2016).
6. Herbst, E. & van Dishoeck, E. F. Complex organic interstellar molecules. *Annu. Rev. Astron. Astrophys.* **47**, 427–480 (2009).
7. Jørgensen, J. K., Belloche, A. & Garrod, R. T. Astrochemistry during the formation of stars. *Annu. Rev. Astron. Astrophys.* **58**, 727–778 (2020).
8. Abramov, O. & Mojzsis, S. J. Abodes for life in carbonaceous asteroids? *Icarus* **213**, 273–279 (2011).
9. Tsuchiyama, A. et al. Discovery of primitive $CO_2$-bearing fluid in an aqueously altered carbonaceous chondrite. *Sci. Adv.* **7**, eabg9707 (2021).
10. Berger, E. L., Zega, T. J., Keller, L. P. & Lauretta, D. S. Evidence for aqueous activity on comet 81P/Wild 2 from sulfide mineral assemblages in Stardust samples and CI chondrites. *Geochim. Cosmochim. Acta* **75**, 3501–3513 (2011).
11. Kaiser, R. I., Stockton, A. M., Kim, Y. S., Jensen, E. C. & Mathies, R. A. On the formation of dipeptides in interstellar model ices. *Astrophys. J.* **765**, 111 (2013).
12. Caro, G. M. M. et al. Amino acids from ultraviolet irradiation of interstellar ice analogues. *Nature* **416**, 403–406 (2002).
13. Potapov, A., Jäger, C., Henning, T., Jonusas, M. & Krim, L. The formation of formaldehyde on interstellar carbonaceous grain analogs by O/H atom addition. *Astrophys. J.* **846**, 131 (2017).
14. Chuang, K.-J. et al. Production of complex organic molecules: H-atom addition versus UV irradiation. *Mon. Not. R. Astron. Soc.* **467**, 2552–2565 (2017).
15. Ioppolo, S. et al. A non-energetic mechanism for glycine formation in the interstellar medium. *Nat. Astron.* **5**, 197–205 (2020).
16. Krasnokutski, S. A., Jäger, C. & Henning, T. Condensation of atomic carbon: possible routes toward glycine. *Astrophys. J.* **889**, 67 (2020).
17. Foden, C. S. et al. Prebiotic synthesis of cysteine peptides that catalyze peptide ligation in neutral water. *Science* **370**, 865–869 (2020).
18. Bujdak, J. & Rode, B. M. Silica, alumina and clay catalyzed peptide bond formation: enhanced efficiency of alumina catalyst. *Orig. Life Evol. Biosph.* **29**, 451–461 (1999).
19. Rodriguez-Garcia, M. et al. Formation of oligopeptides in high yield under simple programmable conditions. *Nat. Commun.* **6**, 8385 (2015).
20. Frenkel-Pinter, M., Samanta, M., Ashkenasy, G. & Leman, L. J. Prebiotic peptides: molecular hubs in the origin of life. *Chem. Rev.* **120**, 4707–4765 (2020).
21. Kitadai, N. & Maruyama, S. Origins of building blocks of life: a review. *Geosci. Front.* **9**, 1117–1153 (2018).
22. Steele, B. A., Goldman, N., Kuo, I. F. W. & Kroonblawd, M. P. Mechanochemical synthesis of glycine oligomers in a virtual rotational diamond anvil cell. *Chem. Sci.* **11**, 7760–7771 (2020).
23. Krasnokutski, S. A. Did life originate from low-temperature areas of the Universe? *Low Temp. Phys.* **47**, 199–205 (2021).
24. Lavallo, V., Canac, Y., Donnadieu, B., Schoeller, W. W. & Bertrand, G. CO fixation to stable acyclic and cyclic alkyl amino carbenes: stable amino ketenes with a small HOMO–LUMO gap. *Angew. Chem. Int. Ed.* **45**, 3488–3491 (2006).
25. Badawi, H. M. Structural stability and vibrational analysis of aminoethylene $CH_2$=CH-$NH_2$ and aminoketene O=C=CH-$NH_2$. *J. Mol. Struct.* **726**, 253–260 (2005).
26. Henning, T. K. & Krasnokutski, S. A. Experimental characterization of the energetics of low-temperature surface reactions. *Nat. Astron.* **3**, 568–573 (2019).
27. Ali, M. F. B. & Abdel-aal, F. A. M. In situ polymerization and FT-IR characterization of poly-glycine on pencil graphite electrode for sensitive determination of anti-emetic drug, granisetron in injections and human plasma. *RSC Adv.* **9**, 4325–4335 (2019).
28. Taga, K. et al. FT-IR spectra of glycine oligomers. *Vib. Spectrosc.* **14**, 143–146 (1997).
29. Olsen, J. V. et al. Higher-energy C-trap dissociation for peptide modification analysis. *Nat. Methods* **4**, 709–712 (2007).
30. Öberg, K. I. Photochemistry and astrochemistry: photochemical pathways to interstellar complex organic molecules. *Chem. Rev.* **116**, 9631–9663 (2016).
31. Snow, T. P. & Witt, A. N. The interstellar carbon budget and the role of carbon in dust and large molecules. *Science* **270**, 1455–1460 (1995).
32. Beuther, H. et al. Carbon in different phases ([CII], [CI], and CO) in infrared dark clouds: Cloud formation signatures and carbon gas fractions. *Astron. Astrophys.* **571**, A53 (2014).
33. Hocuk, S. et al. Parameterizing the interstellar dust temperature. *Astron. Astrophys.* **604**, A58 (2017).
34. Krasnokutski, S. A. et al. Low-temperature condensation of carbon. *Astrophys. J.* **847**, 89 (2017).
35. Pierazzo, E. & Chyba, C. F. Amino acid survival in large cometary impacts. *Meteorit. Planet. Sci.* **34**, 909–918 (1999).
36. Todd, Z. R. & Oberg, K. I. Cometary delivery of hydrogen cyanide to the early Earth. *Astrobiology* **20**, 1109–1120 (2020).
37. Lange, J. et al. A novel proteomics-based strategy for the investigation of peptide sequences in extraterrestrial samples. *J. Proteome Res.* **20**, 1444–1450 (2021).
38. Shimoyama, A. & Ogasawara, R. Dipeptides and diketopiperazines in the Yamato-791198 and Murchison carbonaceous chondrites. *Orig. Life Evol. Biosph.* **32**, 165–179 (2002).
39. Gorokhova, I. V., Chinarev, A. A., Tuzikov, A. B., Tsygankova, S. V. & Bovin, N. V. Spontaneous and promoted association of linear oligoglycines. *Russ. J. Bioorg. Chem.* **32**, 420–428 (2006).
40. Potapov, A., Theule, P., Jäger, C. & Henning, T. Evidence of surface catalytic effect on cosmic dust grain analogs: the ammonia and carbon dioxide surface reaction. *Astrophys. J. Lett.* **878**, L20 (2019).
41. Krasnokutski, S. A. & Huisken, F. A simple and clean source of low-energy atomic carbon. *Appl. Phys. Lett.* **105**, 113506 (2014).
42. Krasnokutski, S. A. et al. Fullerene oligomers and polymers as carriers of unidentified IR emission bands. *Astrophys. J.* **874**, 149 (2019).
43. Frisch, M. J. et al. *Gaussian 16 Rev. C.01* (Gaussian, 2016).
44. Roepstorff, P. & Fohlman, J. Proposal for a common nomenclature for sequence ions in mass-spectra of peptides. *Biomed. Mass Spectrom.* **11**, 601–601 (1984).

## Acknowledgements

This work was financially supported by the Max Planck Institute for Astronomy (MPIA). S.A.K. acknowledges support from Deutsche Forschungsgemeinschaft DFG (grant no. KR 3995/4-1). Th.H. acknowledges support from the European Research Council under the Horizon 2020 Framework programme via the ERC Advanced Grant Origins 83 24 28.

## Author contributions

S.A.K. initiated and led the project. S.A.K. and K.-J.C. recorded and analysed the experimental data. S.A.K. performed the computations and wrote the manuscript. N.U. performed ex situ mass spectrometry. Th.H. and C.J. participated in data interpretation and discussion. All authors contributed to the writing of the manuscript.

## Funding

Open access funding provided by Max Planck Society

## Competing interests

The authors declare no competing interests.

## Additional information

**Extended data** is available for this paper at https://doi.org/10.1038/s41550-021-01577-9.

**Supplementary information** The online version contains supplementary material available at https://doi.org/10.1038/s41550-021-01577-9.

**Correspondence and requests for materials** should be addressed to S. A. Krasnokutski.

**Peer review information** *Nature Astronomy* thanks Cornelia Meinert and the other, anonymous, reviewer(s) for their contribution to the peer review of this work.

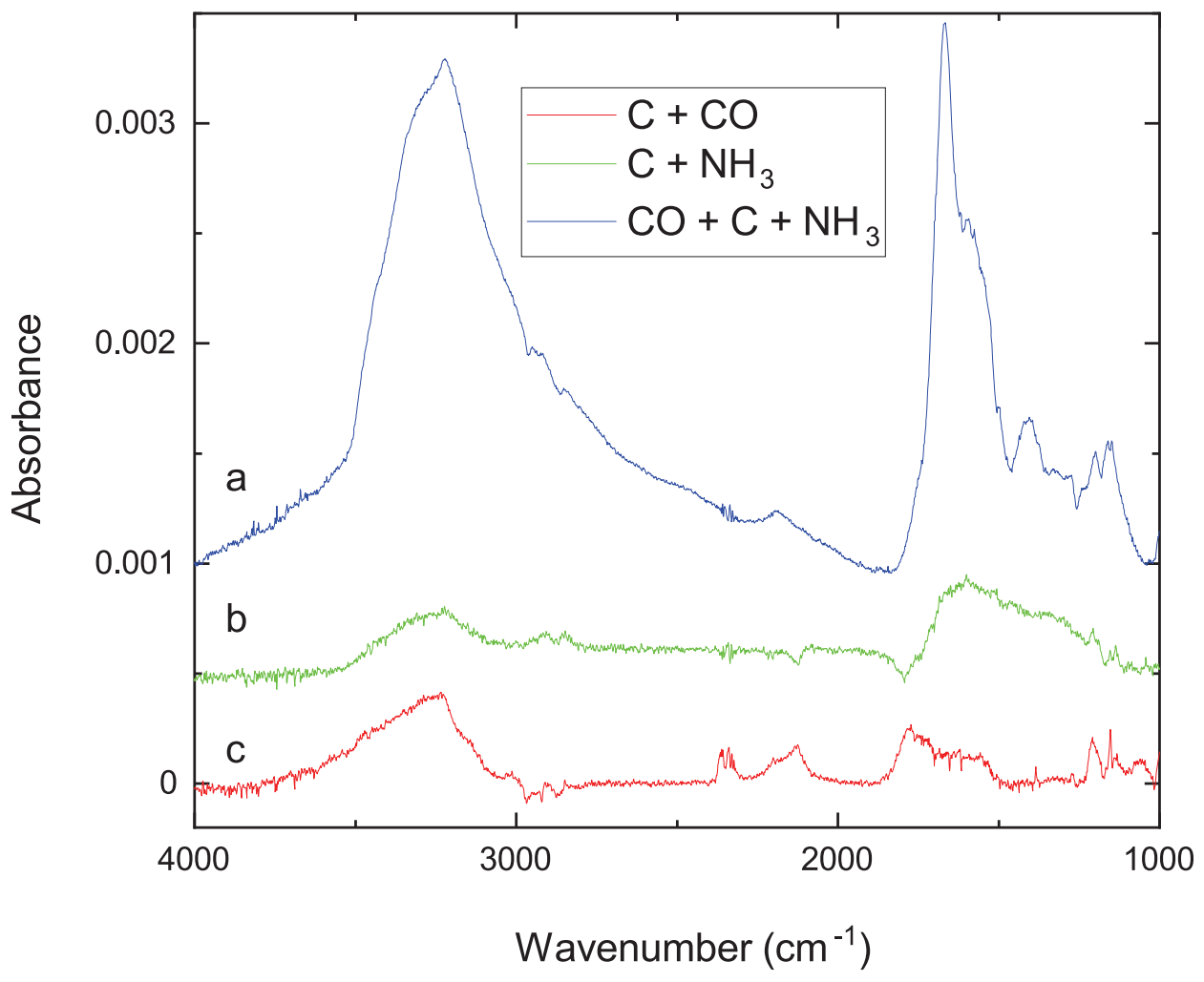

**Extended Data Fig. 1 | The IR absorption spectra of 300 K residues.** a - C + CO + $NH_3$, b - C + $NH_3$, and c - C + CO control experiments.

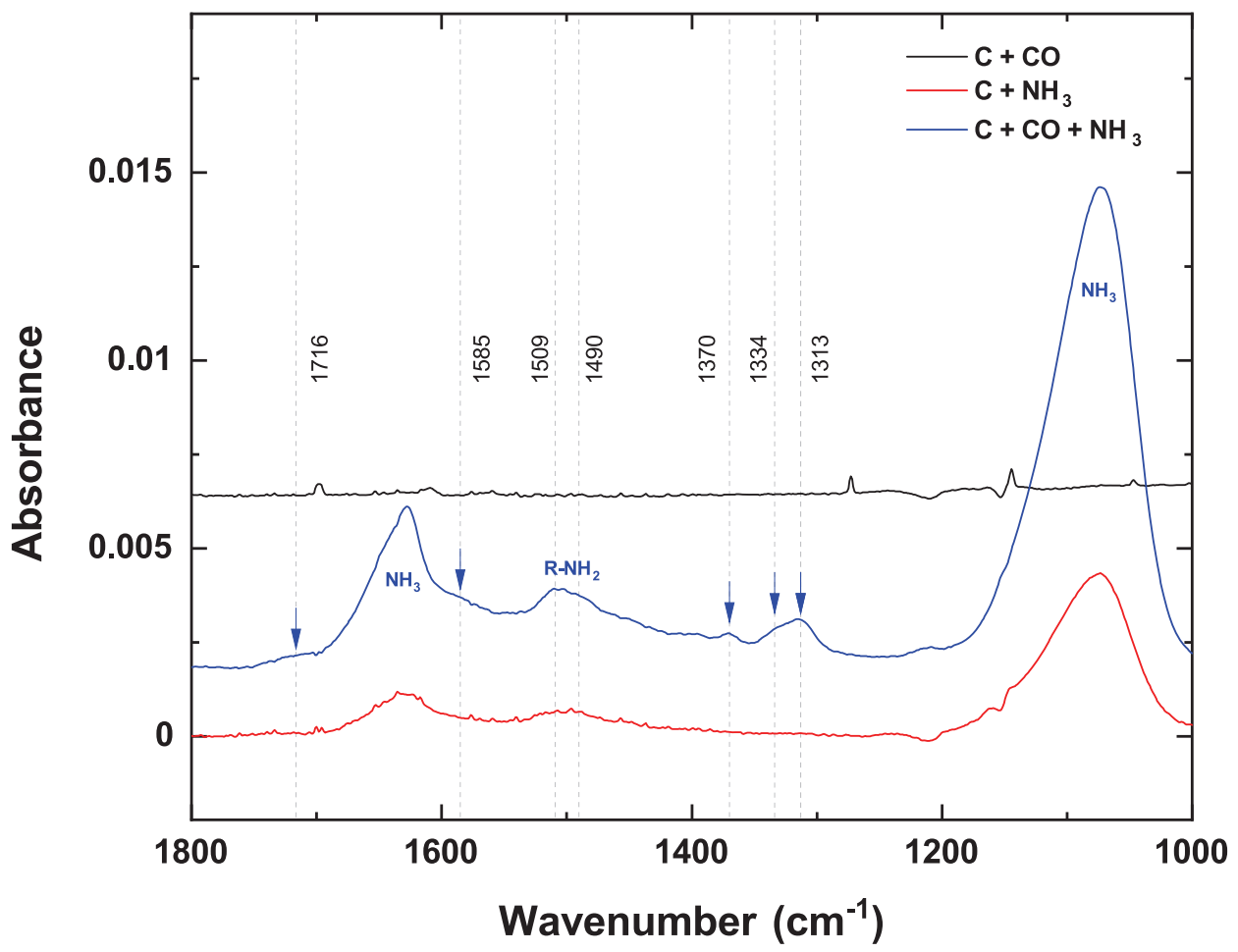

**Extended Data Fig. 2 | The comparison of IR absorption spectra in the range of vibrations of CO and $NH_n$ groups of the materials produced by codeposition of different reactants on the substrate at 10 K.** The bands that are observed only when three reactants are deposited are marked by arrows.

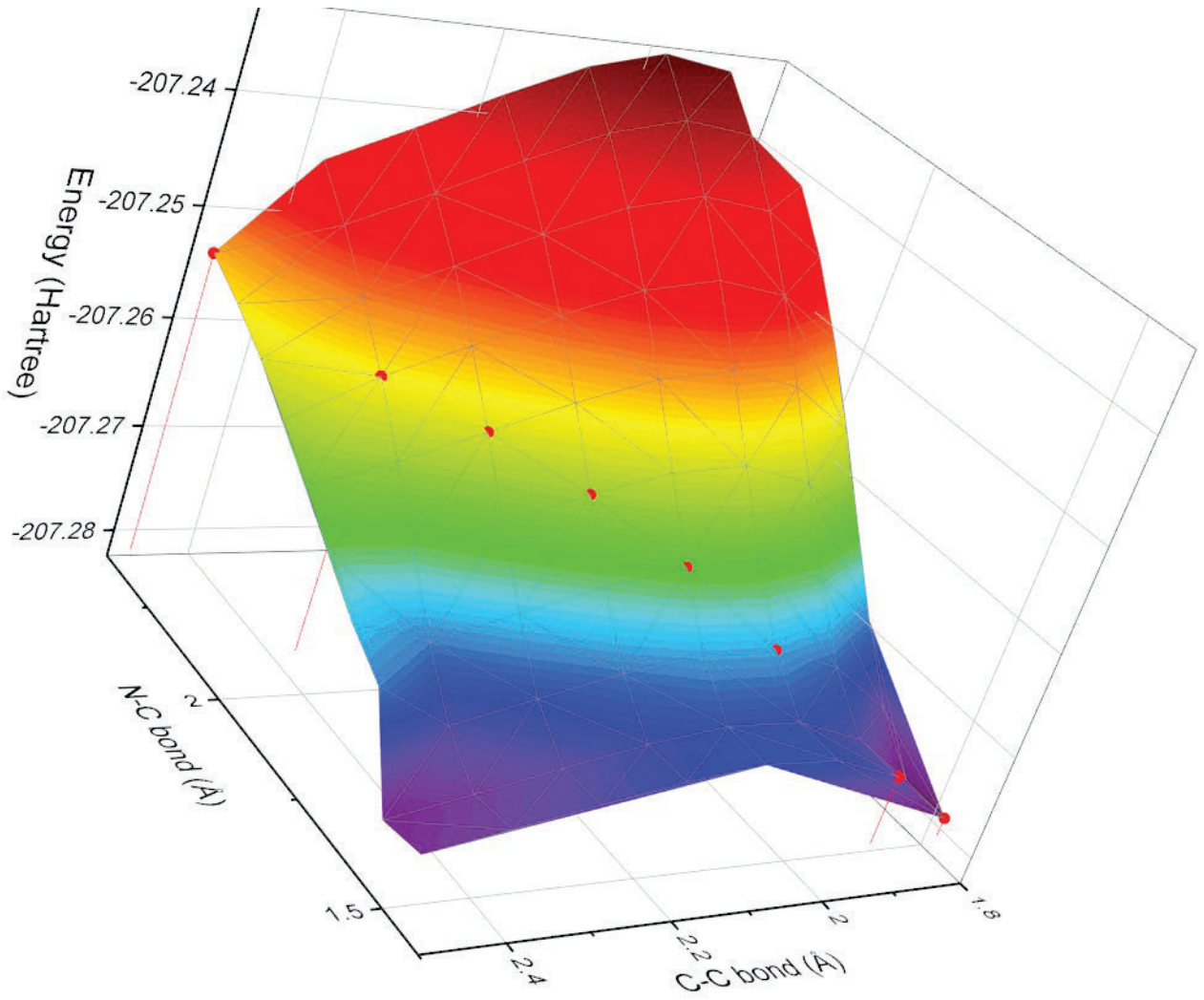

**Extended Data Fig. 3 | Potential energy surface for the formation of the first energy well structure in Fig. 2.** The color of the surface is linked to the energy scale. Big red data points show the barrierless pathway for the formation of the $H_3NCCO$ molecule, while the most probable barrierless path leads to the left minimum and results in the formation of the $H_3NC + CO$ complex, which corresponds to the first energy well structure in Fig. 2. The formation of the $H_3NCCO$ molecule does not notably change the chemical pathway as shown in Fig. 2. In this case, the addition of CO molecule takes place one step earlier, which has very little influence on the energy barrier for the proton transfer.

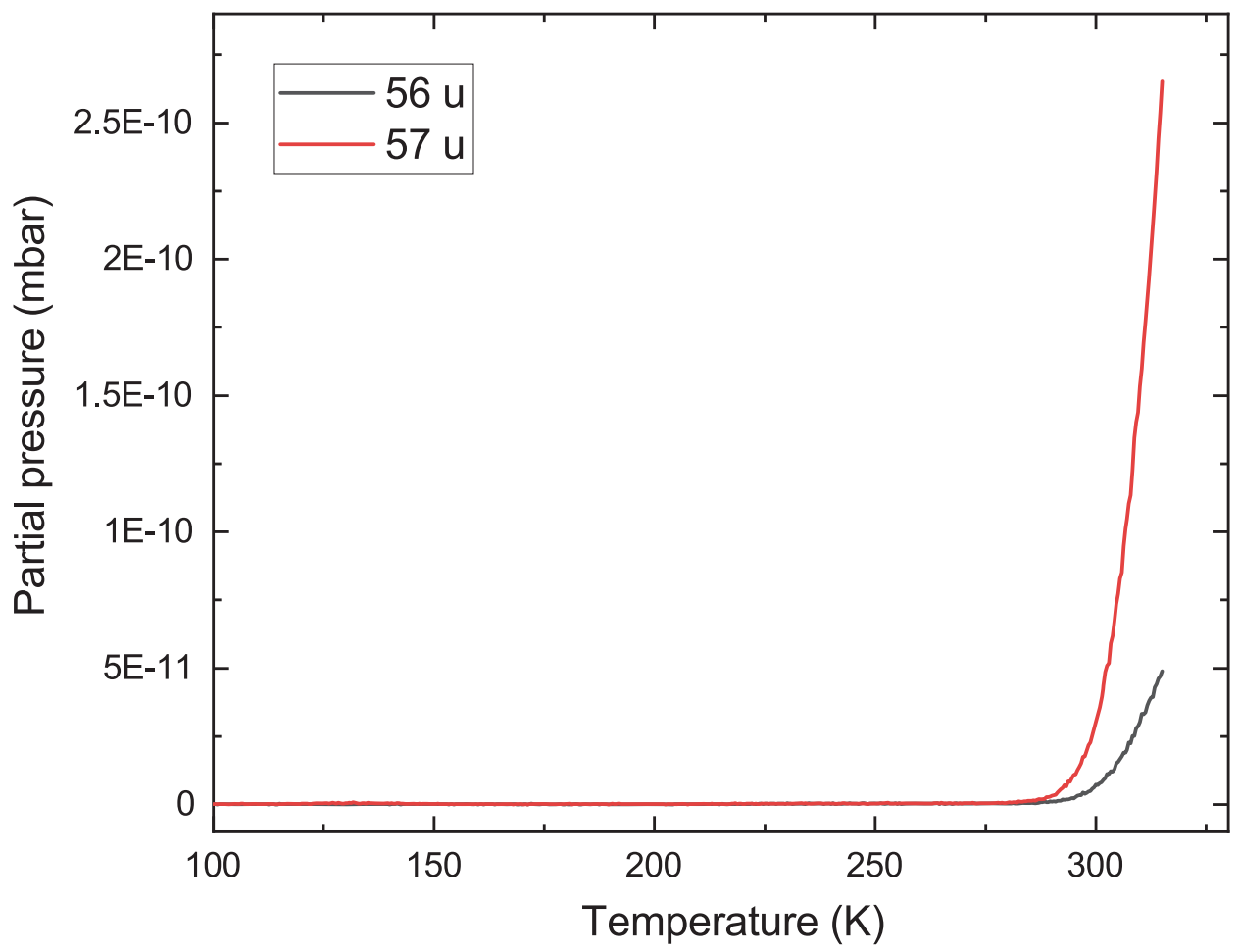

**Extended Data Fig. 4 | The TPD curves measured by quadrupole mass spectrometer monitoring the mass 57 u, corresponding to the $NH_2CH{=}C{=}O$ molecule, and its expected fragment on the mass 56 u.** The partial pressure is given around ionizer of QMS considering the same ionization cross section as for nitrogen.

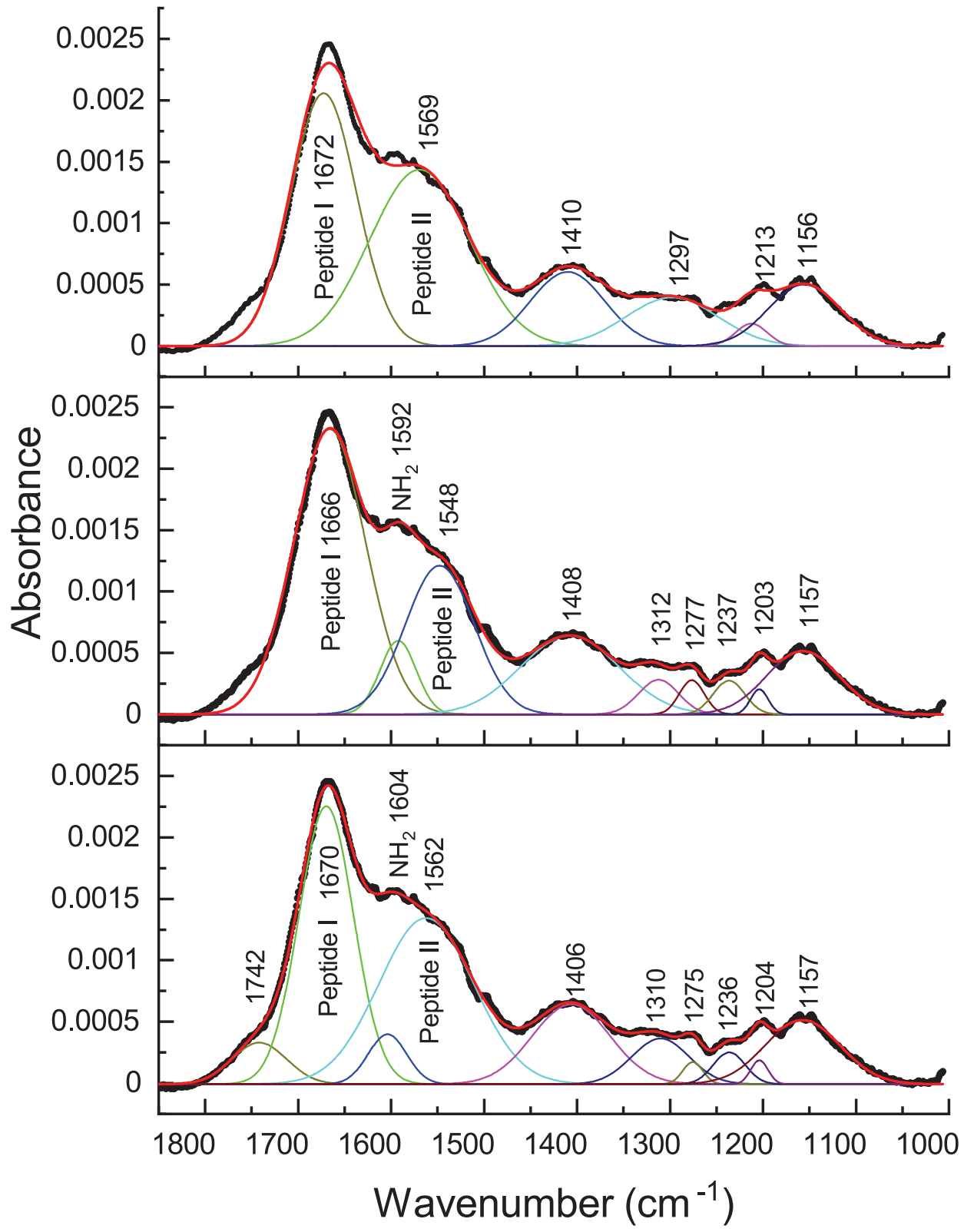

**Extended Data Fig. 5 | The deconvolution of IR spectra of R300K in the range of the peptide bond vibrations.** The dots show the experimental spectrum, the individual Gaussian peaks used for the fit are displayed with thin lines of individual colors, and the cumulative fits are shown by bolder red lines. To demonstrate that there are several possibilities to fit the IR spectrum of R300K, we are providing several alternative deconvolutions of the experimental spectrum. The fitting was performed using Origin software without fixation of any band parameter. Both band positions and widths were optimized to get the closest match to the experimental spectrum. The peptide II band's position varies depending on whether the absorption peak marked as $NH_2$ (1592 or 1604 cm$^{-1}$) is included in the fit. Our calculations predict the presence of a notably intense $NH_2$ scissor vibration band between peptide I and peptide II bands in the spectra of short peptides terminated by amino groups on both sides. Moreover, the match between the experimental spectrum and the fit is much better when the $NH_2$ absorption band is included. Therefore, the deconvolution shown in lower frames seems to be more accurate.

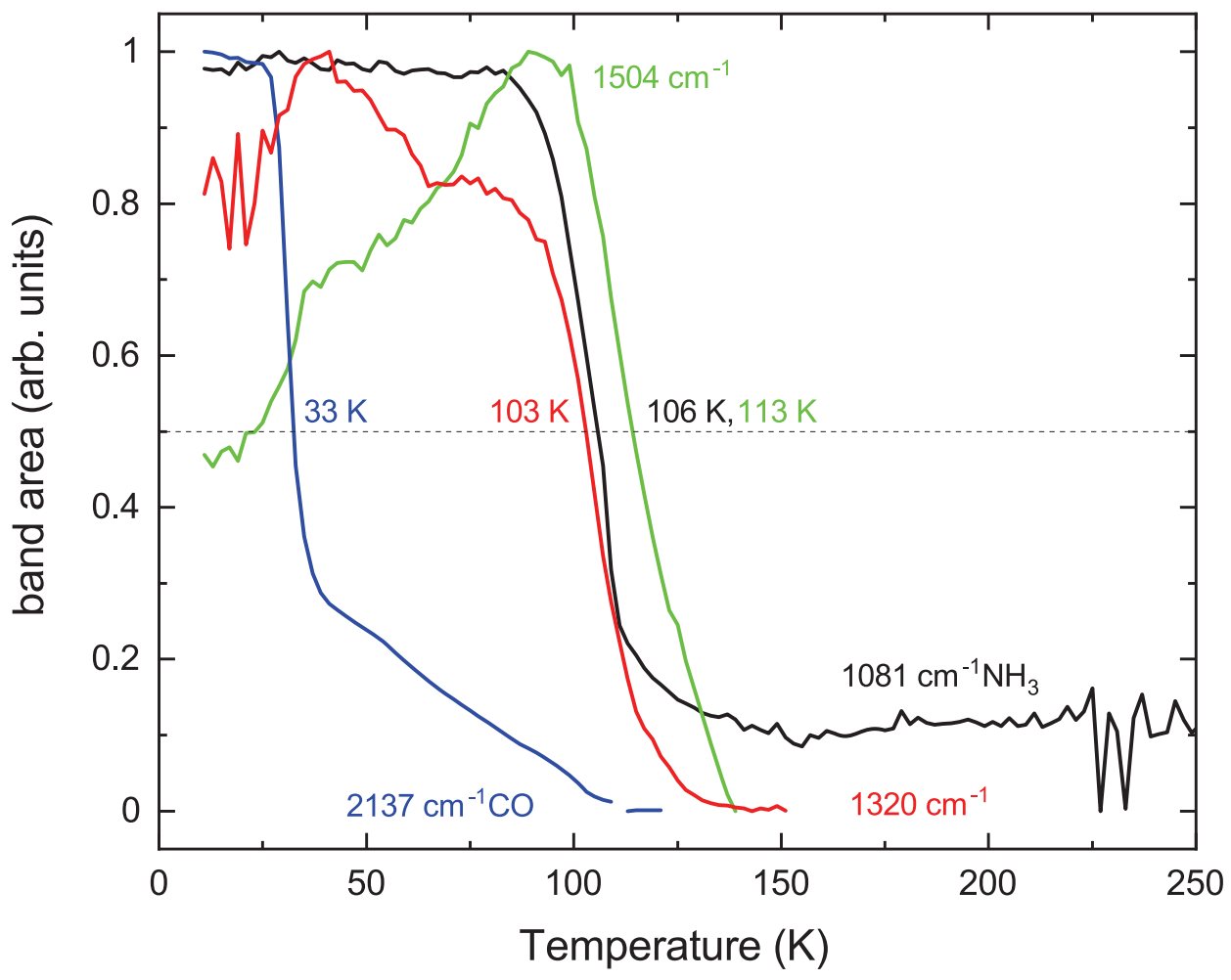

**Extended Data Fig. 6 | The evolution of the intensities of IR absorption bands during the temperature rise.** The curves show the band intensities. The peak positions of the bands and the assignment (if available) are given in the figure. The horizontal dashed line shows the 50% intensity. The temperatures of its crossing with the TPD curves are given. The band at 1081 cm$^{-1}$ marked as $NH_3$ is expected to have a contribution from the solid residue at higher temperatures.

| MASS | COUNTS | ASSIGNMENT | |
|---|---|---|---|
| $NH_2$ $CH_2$ ends | | | |
| 201.0983 | 219568 | $Gl_3$ | $C_7H_{13}N_4O_3$ |
| 258.1195 | 165487 | $Gl_4$ | |
| 315.1412 | 105818 | $Gl_5$ | |
| 372.1627 | 56078 | $Gl_6$ | |
| 429.1842 | 21739 | $Gl_7$ | |
| 486.2057 | 15781 | $Gl_8$ | |
| $NH_2$ $NH_2$ ends | | | |
| 189.098 | 1388446 | $Gl_3$ | $C_6H_{13}N_4O_3$ |
| 246.1194 | 509908 | $Gl_4$ | |
| 303.141 | 121422 | $Gl_5$ | |
| 360.1627 | 50370 | $Gl_6$ | |
| 417.1842 | 18616 | $Gl_7$ | |
| 474.2057 | 11917 | $Gl_8$ | |
| Exact polymer | | | |
| 172.0714 | 362672 | $Gl_3$ | $C_6H_{10}N_3O_3$ |
| 229.093 | 175596 | $Gl_4$ | |
| 286.114 | 93262 | $Gl_5$ | |
| 343.136 | 55089 | $Gl_6$ | |
| 400.157 | 35490 | $Gl_7$ | |
| 457.178 | 16591 | $Gl_8$ | |
| Exact polymer + $H_2$ | | | |
| 174.0871 | 367120 | $Gl_3$ | $C_6H_{12}N_3O_3$ |
| 231.108 | 262322 | $Gl_4$ | |
| 288.13 | 119467 | $Gl_5$ | |
| 345.151 | 65474 | $Gl_6$ | |
| 402.173 | 28327 | $Gl_7$ | |
| 459.194 | 14628 | $Gl_8$ | |
| Exact polymer + CH | | | |
| 184.0715 | 457345 | $Gl_3$ | $C_7H_{10}N_3O_3$ |
| 241.0929 | 250906 | $Gl_4$ | |
| 298.114 | 87861 | $Gl_5$ | |
| 355.136 | 49995 | $Gl_6$ | |
| 412.157 | 25223 | $Gl_7$ | |
| 469.179 | 13877 | $Gl_8$ | |
| Exact polymer + $C_2H_3$ | | | |
| 198.0871 | 238076 | $Gl_3$ | $C_8H_{12}N_3O_3$ |
| 255.108 | 97056 | $Gl_4$ | |
| 312.13 | 77725 | $Gl_5$ | |
| 369.151 | 44642 | $Gl_6$ | |
| 426.173 | 17160 | $Gl_7$ | |
| $NH_2$ $NH_2$ ends + $C_2$ | | | |
| 156.0766 | 383597 | $Gl_2$ | $C_6H_{10}N_3O_2$ |
| 213.098 | 295559 | $Gl_3$ | |
| 270.119 | 128775 | $Gl_4$ | |
| 327.141 | 80091 | $Gl_5$ | |
| 384.162 | 30402 | $Gl_6$ | |
| $NH_2$ $NH_2$ ends + $C_3H$ | | | |
| 168.0766 | 239784 | $Gl_2$ | $C_7H_{10}N_3O_2$ |
| 225.0981 | 259786 | $Gl_3$ | |
| 282.119 | 112662 | $Gl_4$ | |
| 339.141 | 60541 | $Gl_5$ | |
| 396.162 | 37179 | $Gl_6$ | |
| 453.183 | 15029 | $Gl_7$ | |

| MASS | COUNTS | ASSIGNMENT | |
|---|---|---|---|
| $NH_2$ $NH_2$ ends + $H_2$ | | | |
| 191.114 | 13708 | $Gl_3$ | $C_6H_{15}N_4O_3$ |
| 248.135 | 82003 | $Gl_4$ | |
| 305.1567 | 339660 | $Gl_5$ | |
| 362.178 | 44675 | $Gl_6$ | |
| 419.199 | 13724 | $Gl_7$ | |
| $NH_2$ $NH_2$ ends, -O +C | | | |
| 185.1031 | 657156 | $Gl_3$ | $C_7H_{13}N_4O_2$ |
| 242.1245 | 337857 | $Gl_4$ | |
| 299.146 | 123962 | $Gl_5$ | |
| 356.167 | 60163 | $Gl_6$ | |
| 413.188 | 34643 | $Gl_7$ | |
| 470.211 | 15248 | $Gl_8$ | |
| $NH_2$ $NH_2$ ends, -O | | | |
| 173.1031 | 378772 | $Gl_3$ | $C_6H_{13}N_4O_2$ |
| 230.125 | 138257 | $Gl_4$ | |
| 287.146 | 115564 | $Gl_5$ | |
| 344.167 | 33097 | $Gl_6$ | |
| 401.189 | 18335 | $Gl_7$ | |
| $NH_2$ $NH_2$ ends, -O +$CH_2$ | | | |
| 183.0874 | 379942 | $Gl_3$ | $C_7H_{11}N_4O_2$ |
| 240.109 | 148396 | $Gl_4$ | |
| 297.13 | 83123 | $Gl_5$ | |
| 354.152 | 31393 | $Gl_6$ | |
| 411.173 | 13063 | $Gl_7$ | |
| - | | | |
| 157.0606 | 339455 | | $C_6H_9N_2O_3$ |
| 214.082 | 117609 | | |
| 271.151 | 54266 | | |
| 328.125 | 34221 | | |
| 385.147 | 27692 | | |
| - | | | |
| 253.0904 | 327530 | | $C_7H_{15}N_3O_7$ |
| 310.112 | 45682 | | |
| 367.133 | 7652 | | |
| - | | | |
| 204.109 | 20844 | | $C_6H_{14}N_5O_3$ |
| 261.1305 | 277321 | | |
| 318.152 | 18823 | | |
| 375.173 | 15283 | | |
| $NH_2$ $CH_2$ ends + Na | | | |
| 196.069 | 733000 | $Gl_3$ | $C_6H_{11}N_3O_3$ Na |
| 253.0904 | 327530 | $Gl_4$ | |
| 310.112 | 45681 | $Gl_5$ | |
| - | | | |
| 154.0585 | 426849 | | $C_3H_{10}N_2O_5$ |
| 211.08 | 285845 | | |
| 268.101 | 100122 | | |
| 325.123 | 22401 | | |

**Extended Data Fig. 7 | The table of mass peaks found in series in the ex situ analysis of the 300 K residue.** Masses correspond to protonated species $[\text{analyte+H}]^+$ as detected in experiment, while assignment is given for the analyte.